\documentclass[a4paper, 11pt, fleqn]{article}
\usepackage{longtable}
\usepackage{graphicx}
\usepackage{amssymb,amsmath,epsfig,array}

\begin{document}

\title{\bf{Second Outburst of the Yellow Symbiotic Star\\ LT Delphini}}

\author{\it{N.P. Ikonnikova\footnote{E-mail: ikonnikova@gmail.com},
G.V. Komissarova, V.P. Arkhipova}}

\date{Sternberg Astronomical Institute,
Moscow State University, Universitetskii pr. 13, Moscow, 119992
Russia}

\renewcommand{\abstractname}{ }

\maketitle

\begin{abstract}

We present the results of our photoelectric $UBV$ observations of
the yellow symbiotic star LT~Del over 2010--2018. The binary
system LT Del, which consists of a bright K giant and a compact
hot star with a temperature of $\sim$100 000 K, has an orbital
period of 476 days. In 2017 the variable experienced a second
low-amplitude ($\Delta V\sim 0.^{m}7$) outburst in the history of
its studies whose maximum occurred at an orbital phase of
0.15$\pm$0.05. The outburst duration was $\sim$60 days. The $B-V$
and $U-B$ colors in the outburst became noticeably bluer. A
difference in the photometric behavior of the star in the 1994 and
2017 outburst has been detected. In the orbital cycle preceding
the 2017 outburst a secondary minimum with a depth of $0.^{m}15$
and $0.^{m}20$ appeared in the $B$ and $V$ light curves,
respectively, whose cause is discussed. The phase light and color
curves are presented and explained; the position of the star on
the color--color diagram is interpreted. We have estimated the
parameters of the cool and hot components of the system based on
the distance determination from \textit{Gaia} DR2.


\end{abstract}

\section*{Introduction}

The emission-line object He 2-467 from the catalogue of Henize
(1967), which was deemed a planetary nebula for some time after
its discovery, was assigned by Kaler and Lutz (1980) to a small
group of yellow symbiotic stars whose cool components have a
spectral type no later than K5. The binary system He 2-467
consists of a bright G or K giant and a compact hot star with a
temperature of $\sim10^5$ K. The emission-line spectrum is excited
by radiation from the hot star and consists of hydrogen, neutral
and ionized helium lines in the optical band. Based on infrared
(IR) photometry, Allen (1984) classified He 2-467 as an S-type
symbiotic star, a binary system without a circumstellar dust
envelope.

The photometric variability of He 2-467 was discovered by
Arkhipova and Noskova (1985). The authors detected a periodicity
of its brightness variations in the $UBV$ bands with a period of
$\sim$480 days. The shape of the light curves and the growth of
the amplitude from $V$ to $U$ suggested that the variability is
related to the so-called reflection effect, i.e., the heating and
ionization of the upper atmospheric layers of the cool component
facing the hot one by white dwarf radiation. Based on the results
of this paper, the star was included in the catalogue of variable
stars (Samus' et al. 2017) and was designated as LT Del. In
subsequent papers the period was refined and the ephemeris JD(Min)
= 2445930 + 476.$^{d}$0$E$ was adopted based on the 1982--2009
observations (Arkhipova et al. 2011).

In 1994 the star experienced the first outburst that was detected
by Passuello et al. (1994) and studied in detail by Arkhipova et
al. (1995a, 1995b). This event allowed LT Del to be attributed to
classical symbiotic stars. Arkhipova et al. (2011) showed that the
system's cool component does not fill its Roche lobe, the mass
loss occurs in the form of a stellar wind with a rate
$\dot{M}\sim2.5\times10^{-8}$$M_{\odot}$~yr$^{-1}$, and the
outburst activity is associated with slight variations in the rate
of mass loss by the cool component.

In 2017 Munari et al. (2017) reported the detection of a new
outburst. In this paper we present the results of our
photoelectric $UBV$ observations over the period from 2010 to
2018, describe the 2017 outburst, and compare the photometric
behaviors of the star in 1994 and 2017.

\section*{Observations}

Systematic photoelectric observations of LT Del have been
performed with a Zeiss-1 telescope at the Crimean Station of the
Sternberg Astronomical Institute of the Moscow State University
using a photoelectric $UBV$ photometer designed by V.M. Lyutyi
(1971) since 1982. The results of the $UBV$ observations before
2010 were published in Arkhipova and Noskova (1985, 1988) and
Arkhipova et al. (1995a, 2011).

The observations in 2010--2018 were carried out at the same
telescope with the same photometer with a 13$^{\prime \prime}$
aperture. The star BD+19$^{\circ}$4458 with the following
magnitudes was used as a photometric standard: $U=12.^{m}32$,
$B=10.^{m}52$ and $V=8.^{m}98$. We estimate the accuracy of the
observations to be $0.^{m}01-0.^{m}03$ in $V$ and $B$ and at least
$0.^{m}02-0.^{m}04$ in $U$ depending on the brightness level. More
than 100 $UBV$-magnitude estimates were obtained from 2010 to
2018. The results of the observations are presented in
Table~\ref{ubv}.

\begin{center}
\begin{longtable}{@{}cccccc}
\caption{$UBV$ photometry for LT Del in 2010--2018}\\
\label{ubv} \\
\hline
JD&$V$&$B$&$U$&$B-V$&$U-B$\\
\hline
\endfirsthead
\multicolumn{6}{c}%
 {\tablename\ \thetable\  \textit{}} \\ \hline
  JD&$V$&$B$&$U$&$B-V$&$U-B$\\
 \hline \\
\endhead
\multicolumn{6}{r}{\textit{}} \\ \hline
\endfoot
\hline
\endlastfoot

2455305&  13.050&  14.360& 14.210&  1.310& -0.150\\
2455329&  13.040&  14.310& 14.350&  1.270&  0.040\\
2455337&  12.980&  14.380& 14.420&  1.400&  0.040\\
2455385&  13.270&  14.700& 14.790&  1.430&  0.090\\
2455416&  13.270&  15.000& 14.880&  1.730& -0.120\\
2455444&  13.160&  14.800& 15.160&  1.640&  0.360\\
2455448&  13.150&  14.670& 15.120&  1.520&  0.450\\
2455454&  13.070&  14.510& 14.920&  1.440&  0.410\\
2455455&  13.090&  14.540& 14.890&  1.450&  0.350\\
2455502&  13.080&  14.470& 14.560&  1.300&  0.090\\
2455505&  13.100&  14.450& 14.560&  1.350&  0.110\\
2455507&  13.100&  14.460& 14.570&  1.360&  0.110\\
2455508&  13.090&  14.490& 14.540&  1.400&  0.050\\
2455717&  13.020&  14.280& 13.680&  1.260& -0.600\\
2455719&  13.070&  14.270& 13.700&  1.200& -0.570\\
2455743&  12.970&  14.310& 13.790&  1.340& -0.520\\
2455745&  13.050&  14.280& 13.870&  1.230& -0.410\\
2455750&  13.050&  14.310& 13.870&  1.260& -0.440\\
2455751&  13.040&  14.310& 13.880&  1.270& -0.430\\
2455774&  13.030&  14.340& 14.010&  1.310& -0.330\\
2455778&  13.060&  14.360& 14.080&  1.300& -0.280\\
2455781&  13.050&  14.370& 14.030&  1.320& -0.340\\
2455809&  13.110&  14.480& 14.340&  1.370& -0.140\\
2455810&  13.100&  14.440& 14.450&  1.340&  0.010\\
2455831&  13.120&  14.500& 14.530&  1.380&  0.030\\
2455832&  13.130&  14.540& 14.490&  1.410& -0.050\\
2455862&  13.230&  14.620& 14.470&  1.390& -0.150\\
2456040&  13.029&  14.335& 13.959&  1.307& -0.376\\
2456043&  13.038&  14.325& 14.243&  1.287& -0.082\\
2456097&  13.152&  14.337& 13.817&  1.185& -0.520\\
2456182&  13.029&  14.291& 13.693&  1.262& -0.598\\
2456185&  13.013&  14.390& 13.734&  1.377& -0.656\\
2456186&  13.026&  14.313& 13.712&  1.287& -0.601\\
2456216&  13.094&  14.368& 13.911&  1.273& -0.457\\
2456247&  12.944&  14.374& 14.228&  1.430& -0.147\\
2456249&  13.018&  14.374& 14.009&  1.356& -0.365\\
2456431&  13.145&  14.575& 14.672&  1.430&  0.096\\
2456451&  13.135&  14.495& 14.593&  1.360&  0.099\\
2456454&  13.099&  14.472& 14.538&  1.373&  0.066\\
2456480&  13.096&  14.428& 14.361&  1.331& -0.067\\
2456510&  13.089&  14.389& 13.991&  1.300& -0.398\\
2456514&  13.066&  14.383& 14.094&  1.318& -0.290\\
2456534&  13.096&  14.397& 13.975&  1.301& -0.421\\
2456537&  13.091&  14.379& 13.859&  1.288& -0.520\\
2456574&  13.085&  14.362& 13.766&  1.277& -0.597\\
2456591&  13.079&  14.312& 13.720&  1.233& -0.592\\
2456784&  13.165&  14.562& 14.639&  1.397&  0.077\\
2456832&  13.184&  14.710& 15.207&  1.526&  0.496\\
2456837&  13.161&  14.703& 15.366&  1.542&  0.663\\
2456860&  13.202&  14.777& 15.435&  1.575&  0.657\\
2456870&  13.211&  14.754& 15.485&  1.543&  0.731\\
2456893&  13.202&  14.610& 14.892&  1.408&  0.282\\
2456930&  13.058&  14.494& 14.580&  1.436&  0.087\\
2456960&  13.085&  14.353& 14.351&  1.269& -0.002\\
2457006&  13.018&  14.489& 14.011&  1.471& -0.478\\
2457167&  13.017&  14.330& 13.898&  1.313& -0.431\\
2457189&  13.030&  14.371& 14.183&  1.341& -0.188\\
2457217&  13.103&  14.479& 14.349&  1.376& -0.130\\
2457248&  13.127&  14.493& 14.571&  1.367&  0.078\\
2457249&  13.107&  14.536& 14.670&  1.428&  0.134\\
2457250&  13.073&  14.516& 14.617&  1.443&  0.101\\
2457270&  13.058&  14.495& 14.605&  1.437&  0.110\\
2457281&  13.139&  14.535& 14.771&  1.396&  0.236\\
2457301&  13.170&  14.624& 15.122&  1.454&  0.498\\
2457516&  13.045&  14.270& 13.645&  1.224& -0.625\\
2457544&  13.011&  14.196& 13.423&  1.185& -0.773\\
2457545&  13.032&  14.237& 13.419&  1.205& -0.818\\
2457548&  13.036&  14.195& 13.384&  1.159& -0.811\\
2457578&  13.260&  14.338& 13.423&  1.078& -0.915\\
2457600&  13.203&  14.280& 13.132&  1.077& -1.148\\
2457602&  13.259&  14.270& 13.172&  1.011& -1.098\\
2457608&  13.195&  14.318& 13.253&  1.124& -1.066\\
2457612&  13.222&  14.279& 13.219&  1.057& -1.060\\
2457634&  13.070&  14.239& 13.325&  1.169& -0.913\\
2457635&  13.118&  14.254& 13.393&  1.136& -0.861\\
2457639&  13.089&  14.206& 13.500&  1.117& -0.706\\
2457640&  13.127&  14.217& 13.416&  1.090& -0.801\\
2457662&  13.057&  14.223& 13.698&  1.166& -0.525\\
2457714&  13.063&  14.377& 14.222&  1.315& -0.156\\
2457875&  12.478&  13.315& 12.346&  0.837& -0.970\\
2457878&  12.400&  13.261& 12.232&  0.861& -1.029\\
2457905&  12.366&  13.113& 11.971&  0.746& -1.140\\
2457907&  12.368&  13.129& 11.938&  0.761& -1.191\\
2457921&  12.360&  13.157& 11.941&  0.797& -1.216\\
2457930&  12.368&  13.178& 11.962&  0.810& -1.216\\
2457931&  12.421&  13.189& 11.972&  0.769& -1.217\\
2457932&  12.401&  13.187& 11.950&  0.785& -1.237\\
2457934&  12.389&  13.185& 11.989&  0.796& -1.197\\
2457935&  12.436&  13.221& 11.953&  0.784& -1.268\\
2457936&  12.409&  13.210& 11.956&  0.801& -1.254\\
2457950&  12.573&  13.391& 12.145&  0.818& -1.246\\
2457985&  12.858&  13.764& 12.691&  0.906& -1.073\\
2457990&  12.887&  13.769& 12.683&  0.882& -1.086\\
2457994&  12.917&  13.844& 12.771&  0.927& -1.073\\
2458010&  13.040&  13.984& 13.016&  0.944& -0.968\\
2458011&  13.071&  14.019& 13.063&  0.948& -0.955\\
2458013&  13.065&  14.011& 13.068&  0.946& -0.942\\
2458015&  13.104&  14.073& 13.156&  0.970& -0.918\\
2458044&  13.088&  14.047& 13.127&  0.960& -0.921\\
2458253&  13.163&  14.616& 15.115&  1.453&  0.498\\
2458281&  13.176&  14.688& 15.283&  1.512&  0.595\\
2458290&  13.152&  14.657& 15.104&  1.505&  0.448\\
2458309&  13.104&  14.496& 14.744&  1.391&  0.249\\
2458311&  13.129&  14.539& 14.677&  1.411&  0.137\\
2458337&  13.150&  14.548& 14.618&  1.398&  0.070\\
2458338&  13.075&  14.494& 14.706&  1.418&  0.212\\
2458339&  13.122&  14.535& 14.714&  1.414&  0.178\\
2458342&  13.121&  14.521& 14.696&  1.400&  0.175\\
2458344&  13.099&  14.493& 14.671&  1.394&  0.177\\
2458351&  13.090&  14.493& 14.693&  1.403&  0.200\\
2458366&  13.101&  14.435& 14.344&  1.333& -0.091\\
2458374&  13.074&  14.404& 14.196&  1.330& -0.207\\
2458401&  13.045&  14.381& 13.979&  1.336& -0.402\\

\end{longtable}
\end{center}

\section*{Analysis of our $UBV$ photometry}

\subsection*{Light and color curves}

\begin{figure}
 \includegraphics[scale=1.7]{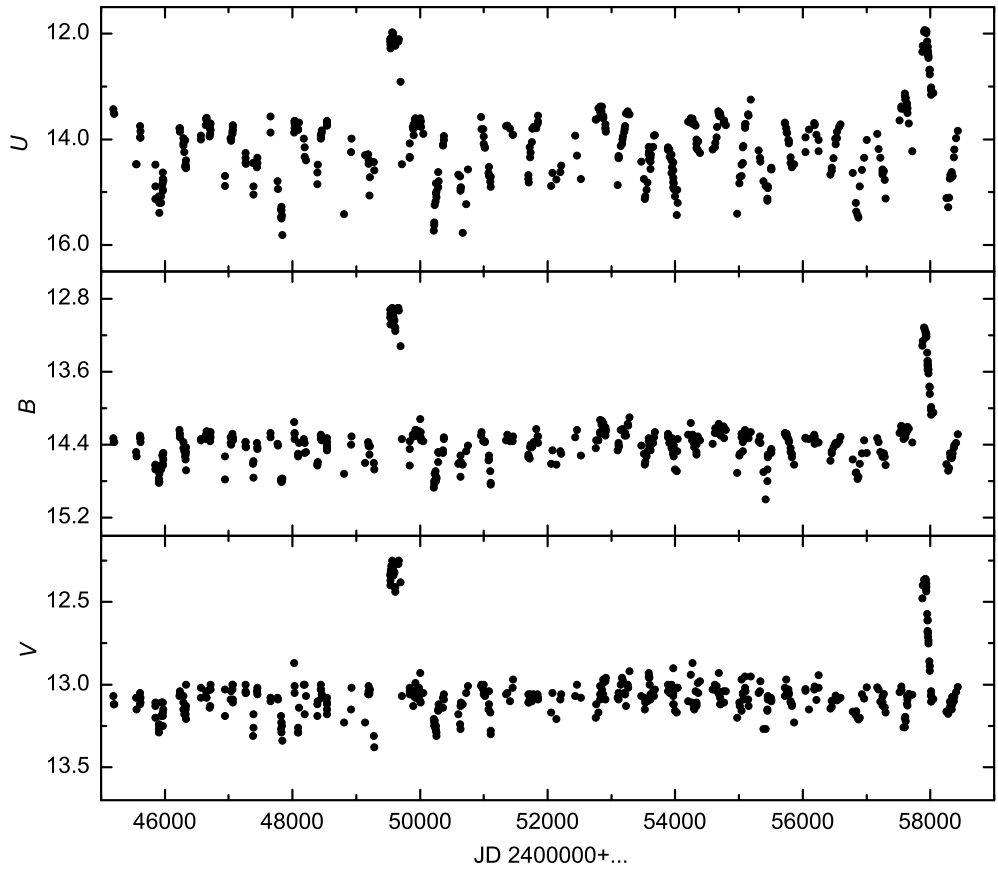}
 \caption{ $UBV$ light curves of LT Del over the period 1982--2018.}
 \label{LC}
\end{figure}

\begin{figure}
 \includegraphics[scale=1.7]{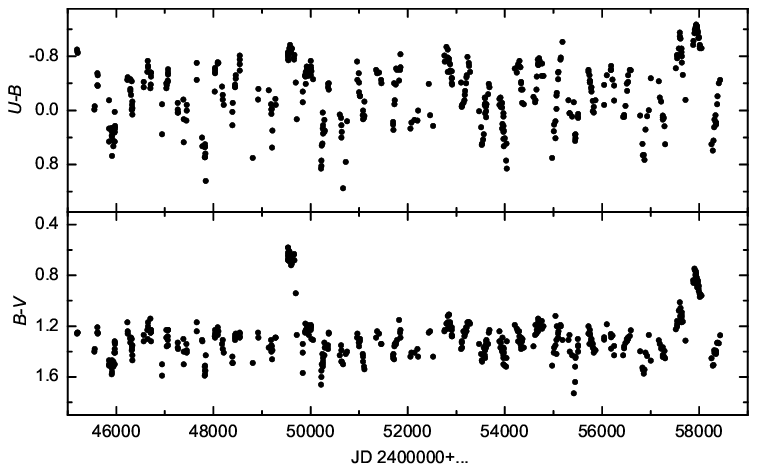}
 \caption{Color curves of LT Del over the period 1982--2018.}
 \label{CI}
\end{figure}

\begin{figure}
 \includegraphics[scale=1.5]{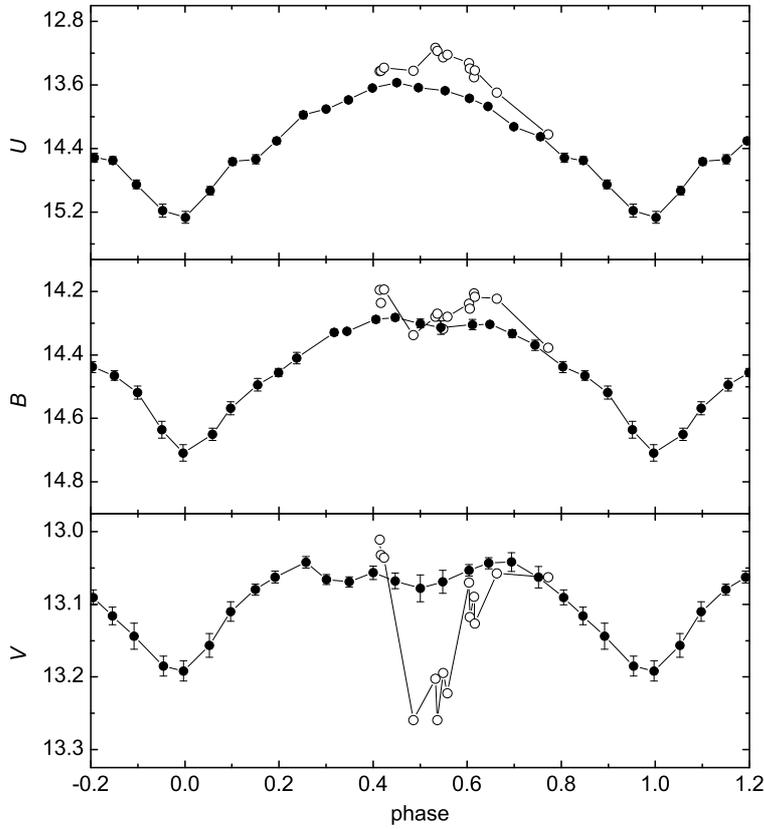}
 \caption{Averaged phase light curves of LT Del in quiescence and observations
 in the pre-outburst stage in 2016 (open circles).}
 \label{phase1}
\end{figure}

Figures~\ref{LC} and ~\ref{CI} show, respectively, the $UBV$ light
and color curves of LT Del over the entire period of its
photoelectric observations in Crimea from 1982 to 2018. In this
time the star experienced two low amplitude outbursts. Outside the
outbursts the photometric variability is related to the changes in
the visibility conditions of the ionization region during the
orbital motion. The amplitude of the brightness variations
increases with decreasing wavelength and is: $\Delta V=0.^{m}2$,
$\Delta B=0.^{m}5$, $\Delta U=1.^{m}8$. The mean magnitudes of the
variable in quiescence are: $V=13.^{m}1$, $B=14.^{m}4$,
$U=14.^{m}2$. Figure~\ref{phase1} presents the phase light curves
averaged with a phase step of 0.05 over the period of observations
from 1982 to 2015, except for the 1994 event. They were computed
with the linear elements: JD(Min) = 2445930 + 476.$^{d}$0$E$ from
Arkhipova et al. (2011). The phase of minimum light 0.0
corresponds to the position when the cool component is in front of
the hot one.

The shape of the orbital $U$ and $B$ light curves for LT Del is
nearly sinusoidal, whereas a fairly broad maximum and a slight
fading at phase 0.5 are observed in the $V$ band. To describe the
shape of the light curve, Skopal (2001) proposed to use the
parameter

\begin{center}
$a=(m(0)-m(0.25))/(m(0)-m(0.5))$, (1)
\end{center}

where $m(0)$, $m(0.25)$, and $m(0.5)$ are the magnitudes at
minimum light, at phase 0.25, and at maximum light, respectively.
For LT~Del we obtained $a_V=1.2$, $a_B=0.8$, $a_U=0.7$. A value of
$a\sim 1$ points to the presence of a secondary minimum in the $V$
light curve. Skopal (2001) discussed in detail the reflection
effect as a possible cause of the observed periodic variability of
a number of symbiotic stars. The reflection effect was shown to
produce a strictly periodic modulation of the light with orbital
phase at $a<0.5$; the amplitude of the brightness variations
should not exceed $0.^{m}1$ in this case. It was concluded that
the observational characteristics of the light curves for
individual symbiotic binaries (including LT Del) cannot be
reproduced by the reflection effect, but are caused by a variation
in the emission measure ($EM$) of the gas component with orbital
phase.

%____________________________fig. 1___________________________________

%_____________________________________________________________________

%____________________________fig. 2___________________________________

%_____________________________________________________________________

In 1994 the star experienced the first outburst in the history of
its observations with amplitudes $\Delta V=0.^{m}8$, $\Delta
B=1.^{m}4$, $\Delta U=1.^{m}9$, during which $B-V$ and $U-B$
became, respectively, noticeably and slightly bluer. The outburst
was observed at orbital phases 0.57--0.90 and lasted $\sim$160
days.

A second such event occurred 23 years later (in 2017). At the very
beginning of the observing season (May 2, 2017) we caught the star
in a bright state. A report on the outburst of LT Del appeared
several days later (Munari et al. 2017). According to Munari et
al. (2017), on May 6, 2017, the star had magnitudes $U=12.^{m}65$,
$B=13.^{m}278$, $V=12.^{m}407$, $R=11.^{m}705$, and $I=11.^{m}135$
and was almost as bright as in the 1994 outburst. All its colors
became bluer.

Over the entire 2017 season we observed the star in some detail
and traced its photometric behavior from May 2 to October 15. The
amplitudes of the new outburst were: $\Delta V=0.^{m}7$, $\Delta
B=1.^{m}3$, and $\Delta U=2.^{m}2$. The $B-V$ and $U-B$ colors
became noticeably bluer. The star stayed in the brightest state
for $\sim$60 days, whereupon it began to fade with a rate of
$0.^{m}015$ per day in the $U$ band. By the end of the 2017
observing season the $V$ brightness of LT Del dropped to its mean
values for the corresponding phases, whereas in the $B$ and $U$
bands the star remained brighter.

Table~\ref{flash} gives some characteristics of the two outbursts
of LT Del: the duration ($\Delta T$), the orbital phases at which
the outburst occurred, the maximum brightness, colors, and
amplitudes of the outbursts. Some difference in the photometric
behavior of the star in the two outbursts should be noted. The
1994 outburst was more prolonged. At an approximately equal
amplitude of the outbursts in the $V$ band the color differed. The
1994 outburst was bluer in $B-V$ and redder in $U-B$.

\begin{figure}
 \includegraphics[scale=1.5]{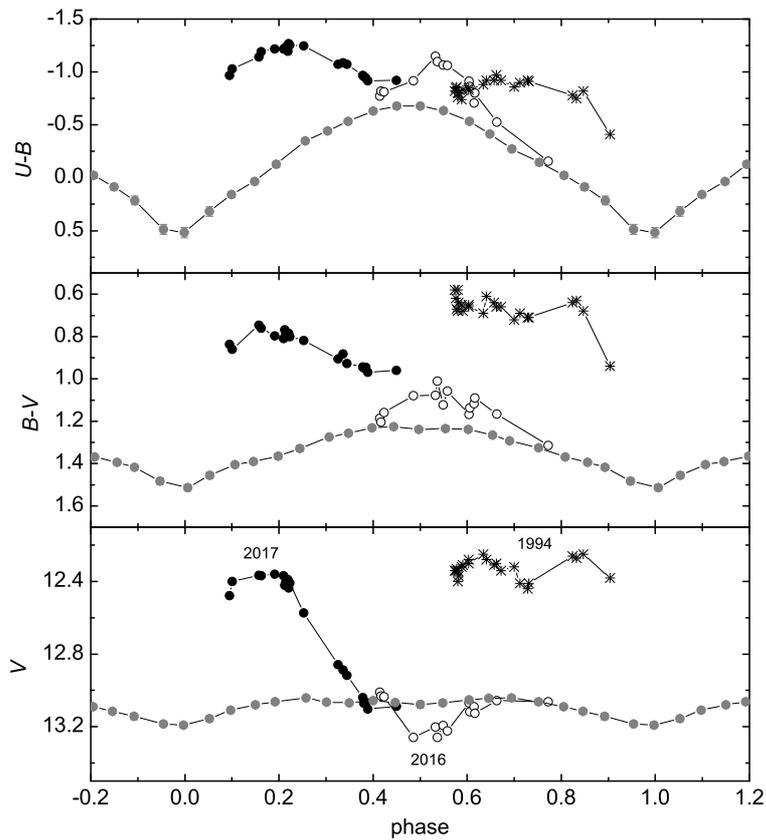}
 \caption{Observations in the 1994 (asterisks) and 2017 (filled circles) outbursts
 as well as in the pre-outburst state in 2016
 (open circles) folded with the orbital period and averaged phase
 light and color curves of LT Del (the gray symbols connected by a line).}
 \label{phase}
\end{figure}

Figure~\ref{phase} show the average phase light and color curves
for quiescence and the position of the star in the 1994 and 2017
outbursts as well as in the preoutburst state in 2016, which
should be discussed in more detail.

In 2016 we carried out the observations from May 7 to November 21
at orbital phases 0.4--0.8. The open circles in the phase curves
(Figs.~\ref{phase1} and ~\ref{phase}) indicate the 2016
observations. In 2016 the star is seen to have deviated noticeably
from the average state. In the $U$ band the star became brighter
by $0.^{m}2-0.^{m}5$, depending on the phase; in the $B$ band a
minor (about $0.^{m}15$) fading was observed against the
background of a general slight brightening at phases 0.4--0.6. A
deep secondary minimum compared to the average level appeared at
these phases in the $V$ band. In 2016 the depth of the secondary
minimum in the $V$ band was $0.^{m}2$ and exceeded the depth of
the primary minimum. The $U-B$ and $B-V$ colors in 2016 became
bluer than those in quiescence at the same orbital phases.

%____________________________fig. 3___________________________________

%______________________________________________________________________

\begin{table}
 \caption{The 1994 and 2017 outbursts}
 \label{flash}

 \begin{tabular}{ccccccccccc}
  \hline
  Year&$\Delta T$, d&Phases&$V$&$B$&$U$&$B-V$&$U-B$&$\Delta V$&$\Delta B$&$\Delta U$\\
    \hline
  1994&$\sim$157&0.57-0.90&12.25&12.90&11.98&0.61&--0.92&$0.8$& $1.4$& $1.9$\\
  2017&$\sim$60&0.10-0.20&12.35&13.12&11.94&0.76&--1.25&$0.7$& $1.3$& $2.2$\\
  \hline

\hline
 \end{tabular}

\end{table}

%____________________________fig. 4___________________________________

%_____________________________________________________________________

\subsection*{$U-B, B-V$ color--color diagram}

\begin{center}
\begin{figure}
 \includegraphics[scale=1.5]{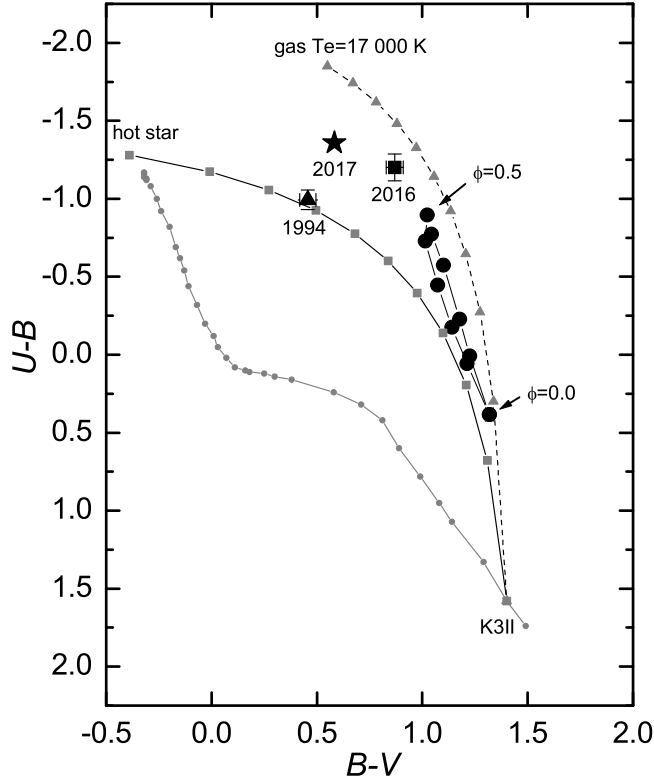}
 \caption{The position of LT Del on the color--color diagram
 corrected for interstellar reddening with $E(B-V)=0.^{m}2$.
 The data for the star in its quiescent state averaged
over a 0.1 phase interval (big circles), the mean colors in the
1994 (triangle) and 2017 (star) outbursts as well as before the
outburst in 2016 (square) are plotted on the diagram. The sequence
of bright giants (Schmidt-Kaler 1982) is represented by the gray
dots connected by a line. The gray triangles connected by a line
are the colors of the total emission from the cool star and the
gas continuum calculated with a 0.1 step in fractions of the
contribution from the individual components to the total emission
in $B$. The gray squares are the same colors for the sum of the
emissions from the hot star and the cool giant.}
 \label{2col}
\end{figure}
\end{center}

Consider the pattern of color variations in LT Del on the $U-B,
B-V$ color--color diagram. We corrected the observed colors for
reddening with $E(B-V)=0.^{m}2$ (Skopal 2005) and plotted the
observations in quiescence averaged over the phase with a 0.1 step
and the data referring to the 1994 and 2017 outbursts and the
pre-outburst state in 2016 on this diagram (Fig.~\ref{2col}). The
positions of the hot star (a blackbody with a temperature $T_{h}$
above 50~000 K), the gas continuum with an electron temperature
$T_{e}=17~000$ K (Skopal 2005), and the sequence of bright giants
(luminosity class II) (Schmidt-Kaler 1982) are also shown on the
graph. For the cool component we adopted the spectral type K3 II
in accordance with the results of Arkhipova et al. (2011). The
figure also plots the colors of the total emission with a variable
fraction of the contribution: (1) the cool component and the gas
continuum; (2) the cool component and the hot star.

The position of LT~Del in the $U-B, B-V$ color--color diagram
changes significantly with orbital phase. In quiescence the
variable moves upward and leftward from phase 0.0 to 0.5
approximately along the line that represents the total emission
from a bright K3 II giant and a variable gas continuum. The $U-B$
range is about $1.^{m}5$ when $B-V$ changes by $0.^{m}3$.
Arkhipova and Noskova (1988) and Arkhipova et al. (1995a)
interpreted this motion as the result of a change in the
visibility of the gas component during the orbital motion of the
binary components.

At minimum light, when the contribution of the hot component
(subdwarf+gas continuum) to the total emission from the system is
no more than 10 \%, the mean observed magnitudes and colors are:
$V=13.^{m}19$, $B=14.^{m}71$, $U=15.^{m}27$, $B-V=1.^{m}52$, and
$U-B=0.^{m}53$. Assuming that the cool star contributes about
90~\%\ in the $B$ band to the total emission from the system, we
obtained the following magnitudes for it: $V=13.^{m}22$,
$B=14.^{m}82$, and $U=16.^{m}53$. In the $V$ band the red giant
accounts for about 97 \%\ of the emission, while in the $U$ band
its contribution is 30 \%.

In the pre-outburst state in 2016 during its deep secondary
minimum LT Del on the color--color diagram shifted relative to the
average position at phases 0.5--0.6 upward and leftward, which can
be explained by a reduction in the contribution of the cool
component due to the ellipticity effect (discussed below) and an
enhancement of the emission-line spectrum of the ionized region.

In the outbursts the variable on the color--color diagram deviated
noticeably from the track of its quiescent state. In 1994 it moved
toward hot stars, which points to a noticeable increase in the
contribution of the hot star to the total emission from the
system. In 1994 the main contributors to the emission were the
erupted subdwarf and the cool star with a flux ratio of 2:1 in the
$B$ band. According to Arkhipova et al. (1995b), the shape of the
emission-line spectrum for LT Del in 1994 underwent no significant
changes compared to its quiescent state and the contribution of
the emission lines to the system's $UBV$ bands was minor.

In the 2017 outburst the ratio of the components in the system's
emission changed differently than in 1994. The influence of the
gas component grew together with a possible increase in the
contribution of the hot star. According to Ikonnikova et al.
(2019), the emission-line fluxes in the optical spectrum of LT Del
increased noticeably (by a factor of 5--10). In the $B$ band the
contribution of the strongest HeII $\lambda$4686 and H$\beta$
lines to the total emission grew by $\sim$13 \%, which led to a
bluing of the $B-V$ color. Munari et al. (2017) reported the
appearance of the high excitation OIII $\lambda$3444 and
$\lambda$3429, OIV $\lambda$3411, [Ne V] $\lambda$3345 and
$\lambda$3427, lines in the outburst, whose contribution to the
$U$ band can be noticeable. Furthermore, the contribution of the
nebular continuum increased. In its brightest state in 2017 LT~Del
had magnitudes $V=12^{m}35$, $B=13.^{m}12$, and $U=11.^{m}94$.
After the subtraction of the cool star with the above magnitudes,
the observed magnitudes of the hot component (star + gas component
and emission lines) in the 2017 outburst were $V=12.^{m}99$,
$B=13.^{m}37$, and $U=12.^{m}10$. Estimating the contribution of
the hot star causes difficulties without additional data on the
ultraviolet (UV) emission from the system.

%____________________________fig. 5___________________________________
%_____________________________________________________________________

\subsection*{Secondary minimum in 2016}

We found that in the orbital cycle before the 2017 outburst a
secondary minimum, which was barely visible in previous years,
clearly manifested itself in the $B$ and $V$ light curves. Both
the ellipticity of the cool component and its eclipse by the hot
star can be responsible for this phenomenon.

Consider the possibility of an eclipse of the cool component by
the hot subdwarf that increased its size. In the $V$ band the main
contributor to the system's emission is the cool star. The depth
of the secondary minimum in 2016, $\Delta V=0.^{m}2$, corresponds
to a decrease in the flux and, accordingly, the area of the disk
of the cool star by a factor of $k$=1.2. In the case of a central
eclipse, when the minimum possible increase in the size of the hot
star is required, from the relation $k=\pi R(\text{cool})^2/(\pi
R(\text{cool})^2-\pi R(\text{hot})^2)$ for $k$=1.2 we obtain
$R(\text{hot})/R(\text{cool})=\sqrt{(k-1)/k}=0.41$, i.e., the
giant and the hot subdwarf must have comparable sizes. As will be
shown below, $R(\text{cool})\sim 35R_{\odot}$, and then
$R(\text{hot})=0.41\times35R_{\odot}\sim 14R_{\odot}$. According
to spectroscopic observations, the temperature of the hot star in
2016 did not decrease; consequently, such a significant increase
in the radius compared to the system's quiescent state, when the
subdwarf has a radius $R(\text{hot})\sim0.26R_{\odot}$, would
entail a manifold increase in the system's luminosity, which was
not observed. It remains to recognize that the secondary minimum
was caused by a manifestation of the giant's ellipticity before
the 2017 outburst. However, this interpretation is also
questioned, as shown in the Section "Distance and Estimation of
Component Parameters"\ .

The phenomenon detected by us together with the model calculations
of light curves will be discussed later.

\section*{Distance and estimation of
component parameters}

Using the color--color diagram and the results of Skopal (2005),
we estimated the magnitudes of the components of the binary system
LT Del.

At the phase of maximum light ($\phi=0.49$) only one UV spectrum
of LT Del was obtained from the IUE satellite on June 14, 1990.
These data together with the photometric observations in a wide
wavelength range, from 0.36 ($U$) to 5.5 ($L$) $\mu$m, from Munari
and Buson (1992) allowed Skopal (2005) to construct the spectral
energy distribution for LT~Del at this orbital phase and to
represent it by three components: a hot star with
$T(\text{hot)}=10^5$ K, a gas continuum with $T_e=17~000$ K, and a
cool component with $T_\text{eff}=4100$ K. Using Fig. 12 from
Skopal (2005), we estimated the magnitudes of the hot star
corrected for reddening with $E(B-V)=0.^{m}2$ to be $V=16.^{m}36$,
$B=15.^{m}99$, and $U=14.^{m}69$. The accuracy of these estimates
is $0.^{m}1$.

At minimum light, when the cool star contributes about 90 \%\ to
the total emission from the system in $B$, we obtained the
following observed magnitudes for it: $V=13.^{m}22$,
$B=14.^{m}82$, $U=16.^{m}53$. The $V$ magnitude of the cool star
corrected for reddening with $E(B-V)=0.^{m}2$ is $V=12.^{m}60$.

One cannot but note the distance determination for LT Del based on
the latest\textit{Gaia} data. A simple passage from
the\textit{Gaia} DR2 parallax $\pi=0.0874\pm0.0536$ mas (Gaia
Collaboration, 2018) to the distance led to an unrealistic value.
Using a more complex procedure allowed Bailer-Jones et al. (2018)
to get the distance $d=5810.6$ pc with the confidence interval
4612--7533 pc. LT Del has the following Galactic coordinates
$l=063.^{\circ}40$ and $b=-12.^{\circ}15$. With its new distance
estimate the z coordinate for the star is $|z|=1223^{+360}_{-240}$
pc.

Given the distance and adopting the dereddened magnitude of the
cool component $V(\text{cool})=12.^{m}60$, we obtained its
absolute magnitude by taking into account the distance errors:
$M_V(\text{cool})=(-1.22^{+0.50}_{-0.56})^{m}$. With this $M_V$
the cool star occupies an intermediate position between a giant of
spectral type K3 with $M_{V}=0.^{m}6$ and a bright K3 giant with
$M_{V}=-2.^{m}5$ (Strai\v{z}ys 1982). Taking the bolometric
correction as a mean between $BC(\text{K3III})=-0.^{m}66$ and
$BC(\text{K3II})=-0.^{m}57$ (Strai\v{z}ys 1982), we obtained the
bolometric absolute magnitude of the cool star
$M_{bol}(\text{cool})\sim-1.^{m}84$ and its luminosity
$L(\text{cool})\sim420L_{\odot}$ or $\log L/L_{\odot}\sim2.62$.
For a star with $T_{\text{eff}}=4400$ K (Pereira et al. 1998) the
radius is $R(\text{cool})\sim35R_\odot$. With the deduced
parameters and given the chemical composition [Fe/H]=--1.1
(Pereira et al. 1998), the position of the cool giant on the
Hertzsprung--Russell diagram is consistent with the evolutionary
track of a star with metallicity $Z$=0.001 and mass
0.93$M_{\odot}$ at the red giant stage (Charbonnel et al. 1996).

We also estimated the luminosity of the hot star in the system LT
Del. As has been shown above, the subdwarf in quiescence has a
magnitude corrected for reddening with $E(B-V)=0.^{m}2$ of
$V(\text{hot})=16.^{m}36$, implying
$M_{V}(\text{hot})=(1.93^{-0.57}_{+0.49})^{m}$ for a distance
$d=5811_{-1199}^{+1722}$ pc. We calculated the bolometric
correction using the equation $BC=27.66-6.84\log T_\text{eff}$
from Vacca et al. (1996). For $T(\text{hot})=10^5$ K (Skopal 2005)
we obtained $BC=-6.^{m}64$. With this $BC$ the hot star has a
bolometric absolute magnitude $M_{bol}(\text{hot})\sim-4.^{m}61$
and a luminosity $L(\text{hot})\sim5400L_{\odot}$ or $\log
L/L_{\odot}\sim3.73$. The radius of the subdwarf is estimated to
be $R \sim 0.26R_{\odot}$. According to the present-day
evolutionary model at late stages of intermediate-mass stars by
Miller Bertolami (2016), the hot subdwarf in the system LT Del
with the above parameters $T_\text{eff}\sim 10^{5}$ K, $\log
L/L_\odot\sim3.73$, and $Z$=0.001 has a mass of $\sim$
0.57$M_\odot$.

With the new mass estimates for the components of the binary
system with an orbital period $P=476^{d}$ under the assumption of
a circular orbit, we estimated the separation between the
components to be 294$R_\odot$, and the Roche lobe size for the
cool giant to be 85$R_\odot$. Consequently, the star with a radius
$R(\text{cool})\sim35R_\odot$ in the system LT Del does not fill
its Roche lobe and its shape is not distorted by the tidal
interaction with the hot component. This casts doubt on the
conclusion about a possible ellipticity of the cool component as
the cause of the secondary minimum in the $V$ light curve in 2016.

\section*{Conclusions}

In 2010--2018 we obtained new photoelectric $UBV$ observations of
the yellow symbiotic star LT Del. Based on our data and the
results of other authors, in particular, the latest distance
determination, we reached the following conclusions:

(1) An outburst with amplitudes $\Delta V=0.^{m}7$, $\Delta
B=1.^{m}3$, and $\Delta U=2.^{m}2$. was shown to have occurred in
the binary system in 2017. The outburst duration was $\sim$60
days. In the outburst the star had considerably bluer colors than
in quiescence.

(2) A comparison of the 1994 and 2017 outbursts showed their
difference. The 1994 event was more prolonged. At an approximately
equal amplitude in the $V$ band, the 1994 outburst was bluer in
$B-V$ and redder in $U-B$ than the 2017 event. This points to a
larger contribution of the gas component in 2017. Given the 2017
spectroscopic data, which showed a noticeable strengthening of
high-excitation lines, it can be hypothesized that in 2017 the
temperature of the hot star rose, while in the 1994 outburst the
compact object became cooler, as shown in Arkhipova et al.
(1995b).

(3) In the pre-outburst state in 2016 secondary minima were
observed in the $B$ and $V$ light curves at orbital phases
0.4--0.6, a reasonable interpretation of which requires additional
knowledge of the binary system.

(4) We estimated the parameters of the cool and hot components of
the system by taking into account the direct distance
determination for the binary system from\textit{Gaia} data. The
cool star was found to be a yellow giant with $L (\text{cool})\sim
420 L_{\odot}$ and $R(\text{cool}) \sim 35R_{\odot}$, while the
hot subdwarf has $L(\text{hot})\sim 5400 L_{\odot}$ and
$R(\text{hot}) \sim 0.26R_{\odot}$. A comparison of the deduced
binary component parameters with theoretical evolutionary tracks
allowed the component masses to be estimated:
$M(\text{cool})\sim0.93M_{\odot}$ and $M(\text{hot})\sim
0.57M_{\odot}$.

\bigskip
REFERENCES
\bigskip

\begin{enumerate}

\item Allen D.A. , Proc. Astron. Soc. Austral. {\bf 5}, 369
(1984).

\item Arkhipova V.P., Esipov V.F., and Ikonnikova N.P., Astron.
Lett. {\bf 21}, 439 (1995b).

\item Arkhipova V.P., Esipov V.F., Ikonnikova N.P., Komissarova
G.V., and Noskova R.I., Astron. Lett. {\bf 37}, 377 (2011).

\item Arkhipova V.P., Ikonnikova N.P., and  Noskova R.I., Astron.
Lett. {\bf 21}, 379 (1995a).

\item Arkhipova V.P. and  Noskova R.I., Astron. Lett. {\bf 11},
706 (1985).

\item Arkhipova V.P. and  Noskova R.I., Astron. Lett. {\bf 14},
445 (1988).

\item Bailer-Jones C.A.L., Rybizki J., Fouesneau M.,  Mantelet G.,
and  Andrae R., Astron. J., 156:58 (2018).

\item Brown, A. G. A., Vallenari, A., Prusti, T. et al. (Gaia
Collab.), Astron. Astrophys., {\bf 616}, 10 (2018).

\item Charbonnel C., Meynet G., Maeder A., and  Schaerer D.,
Astron. Astrophys. Suppl. Ser. {\bf 115}, 339 (1996).

\item Henize K.G., Astrophys. J., Suppl. Ser. {\bf 14}, 125
(1967).

\item Ikonnikova N.P., Burlak M.A.,  Arkhipova V.P., and Esipov
V.F., Astron. Lett. {\bf 45}, 217 (2019).

\item Kaler J.B. and Lutz J.H., PASP {\bf 92}, 81 (1980).

\item Lyutyi V.M., Soobshch. GAISh {\bf 172}, 30 (1971).

\item Miller Bertolami M.M., Astron. Astrophys. {\bf 588}, A25
(2016).

\item Munari U. and Buson L.M., Astron. Astrophys. {\bf 255}, 158
(1992).

\item Munari U., Ochner P., Dallaporta S., and Belligoli R.,
Astronomer's Telegram, 10361 (2017).

\item Passuello R. et al.,  IAUC 6065 (1994).

\item Pereira C.B., Smith V.V., and Cunha K., Astron. J. {\bf
116}, 1997, (1998).

\item Samus' N.N., Kazarovets E.V., Durlevich O.V., Kireeva N.N.,
and Pastukhova E.N., Astron. Rep. 61, 80 (2017).

\item Schmidt-Kaler Th. 1982, in: K. Schaifers and H.H. Voigt
(eds.), Landolt-Bornstein, Numerical Data and Functional
Relationships in Science and Technology (New Series, Group VI,
Vol. 2b), Springer Verlag, Berlin.

\item Skopal A., Astron. and Astrophys. {\bf 366}, 157 (2001).

\item Skopal A., Astron. and Astrophys. {\bf 440}, 995 (2005).

\item Strai\v{z}ys V.L., Stars with Metal Deficite (Mokslas,
Vil'nyus, 1982) [in Russian].

\item W.D. Vacca, C.D. Garmany, and J.M. Shull, Astrophys. J. {\bf
460}, 914 (1996).

\end{enumerate}

\end{document}